\documentstyle[preprint,eqsecnum,aps,epsfig]{revtex}\tighten

\newcommand{\beq}{\begin{equation}}
\newcommand{\beqa}{\begin{eqnarray}}
\newcommand{\eeq}{\end{equation}}
\newcommand{\eeqa}{\end{eqnarray}}

\newcommand{\siml}{\lesssim}
\newcommand{\C}{{\cal C}}

\begin{document}

\draft
\preprint{YITP-98-12, gr-qc/9904054}

\title{
Apparent Horizon Formation and Hoop Conjecture in 
Non-axisymmetric Spaces
}
\author{Takeshi Chiba\footnote{Current address: Department of Physics, 
University of Tokyo, Tokyo 113-0033, Japan}}
\address{
Yukawa Institute for Theoretical Physics, Kyoto University, 
Kyoto 606-01, Japan}

\date{\today}

\maketitle

\begin{abstract}
We investigate the validity of Thorne's hoop conjecture 
in non-axisymmetric spacetimes by examining the formation of apparent
horizons numerically. If spaces have a discrete symmetry about one
axis, we can specify the boundary conditions to determine an apparent
horizon even in non-axisymmetric spaces. 
We implement, for the first time, the ``hoop finder'' in 
non-axisymmetric spaces with a discrete symmetry. We construct
asymptotically flat vacuum solutions at a moment of time symmetry. 
Two cases are examined: black holes distributed on a ring, and black holes 
on a spherical surface. It turns out that calculating ${\cal C}$ is
reduced to solving an ordinary differential equation. We find that
even in non-axisymmetric spaces the existence or nonexistence of
an apparent horizon is consistent with  the inequality: 
${\cal C} \siml 4\pi M$. 
\end{abstract}


\section{introduction}

In 1972, Kip Thorne proposed the hoop conjecture \cite{hc}; 
{\it horizons form when and only when a mass $M$ gets compacted into a
region whose circumference in every direction is 
${\cal C} \siml 4\pi M$.} 
However, it took nearly a decade until 
the studies 
appeared which directly address the conjecture.
This may be because of the vagueness of the statement, i.e.,  
the circumference ${\cal C}$  of the horizon and the mass $M$ of a body
are ambiguously defined. 

For axisymmetric bodies, the definition of ${\cal C}$ is rather simple
as either  the minimum equatorial circumference or the minimum polar 
circumference\cite{nst,fla}. Hence there exist a lot of works both by
numerical and by analytical approaches addressing the hoop conjecture in
the axisymmetric system\cite{nst,st,ahst,cnns}.  

For non-axisymmetric bodies, on the other hand, we didn't have any 
satisfactory definition of ${\cal C}$ until we proposed a
definition using closed geodesics\cite{cnns}. 
In this paper we demonstrate how to calculate ${\cal C}$ 
in numerically generated spacetime without axisymmetry but with a 
discrete symmetry and assess the validity of the hoop conjecture. 
We study the formation of apparent horizons because apparent horizons 
can be identified locally and provide the practical definition 
of a black hole. It is important to emphasize here that event horizons
may clothe the singularities even if apparent horizons do not appear
on the spatial slices considered. The existence of an apparent horizon is a
sufficient condition for the formation of an event horizon\cite{he}.

Unfortunately there exists an obstacle to study the formation of apparent
horizons numerically because we do not have an efficient and robust
method to find apparent horizons in general non-axisymmetric spaces, 
although there has been recent progress\cite{nko,bcsst,3d,gun}. 
This is an important aspect because we are
interested in the existence or nonexistence of an apparent horizon. 
On the other hand, if we allow spaces to have a
discrete symmetry (but not continuous one), we can solve the equation
for the surface of an apparent horizon, which is generally a nonlinear 
elliptic equation, as a boundary value problem\cite{3d}. Then the
numerical treatment of the problem becomes quite similar to that in
the axisymmetric case. 

In this paper, taking advantage of recent progress both in the 
definition of the circumference  and in finding apparent horizons
numerically,  
we study, for the first time, the condition 
for the formation of apparent horizons in the light of the hoop conjecture 
in spaces without axisymmetry but with a discrete symmetry. 
We present 
the momentarily static vacuum configurations. 
We consider two families of configurations as a 
demonstration: black holes distributed on a ring, and black holes 
on a spherical surface. We increase the number of black holes keeping
the total mass constant and explore the existence of a common apparent 
horizon. As for the definition of a mass we simply adopt the 
ADM mass, because of its uniqueness and definiteness at least for the
 initial data on a spacelike hypersurface.

This paper is organized as follows. 
In Sec.2, first we review the method for finding apparent horizons 
in spaces with a discrete symmetry, and then  
we also review a definition of the circumference of a 
body which has been proposed by us, and 
finally we present the results of our 
numerical analysis of the initial data. Section 3 is devoted to
summary.

\section{initial data analysis}

\subsection{Finding Apparent Horizons}

A marginally trapped surface  is a closed 2-surface $S$ where 
the expansion $\Theta $ of  future-directed 
outgoing null vectors $\ell^{\mu}$ normal to it vanishes\cite{he}. 
An apparent horizon is defined as the outer boundary of a connected
component of the trapped region. When the apparent horizon forms, there
always exists the event horizon enclosing it if 
there is no naked singularity and the null convergence condition 
is satisfied\cite{he}. 

Let $s^{\mu}$ be the outward-pointing spacelike unit normal to $S$
and $n^{\mu}$ be the unit normal to a time slice. Then $\ell^{\mu}$
can be written as $\ell^{\mu}=n^{\mu}+s^{\mu}$ and thus
\beq
\Theta=D_{i}s^i+K_{ij}s^is^j-K=0
\label{expansion}
\eeq
on $S$. 

In this paper,  we consider the conformally flat space
for simplicity
\beq
dl^2=\psi^4 f_{ij}dx^idx^j,
\eeq
where $f_{ij}$ denotes the flat metric.  Then Eq.(\ref{expansion}) 
is rewritten as 
\beqa
\Theta=&-&{r\over \psi^2(r^2+r^2_{\theta})^{3/2}}
\Bigl[r_{\theta \theta}+r_{\theta}\cot \theta +{r_{\phi \phi}\over
\sin ^2\theta} -2 r -{3\over r}\left(r_{\theta}^2+{r_{\phi}^2\over
\sin ^2\theta} \right) \nonumber\\
&-& 4\left( {\psi_{r}\over \psi}-{\psi_{\theta}\over \psi}
{r_{\theta}\over r^2}-{\psi_{\phi}\over \psi}{r_{\theta}\over 
r^2 \sin ^2\theta}\right)\left(r^2+r_{\theta}^2+{r_{\phi}^2\over 
\sin^2 \theta}\right)
+{r_{\theta}^2\over r^2}\left( r_{\theta}\cot \theta + {r_{\phi 
\phi}\over \sin ^2 \theta}\right)\nonumber\\
&-&2{r_{\theta }r_{\phi}\over r^2 \sin ^2 \theta}\left(r_{\theta \phi}-
r_{\phi} \cot \theta \right)+{r_{\theta\theta}r_{\phi}^2\over r^2 
\sin ^2 \theta}\Bigr]
+K_{ij}s^is^j-K=0,
\label{ah}
\eeqa
where $r(\theta,\phi)$ parameterizes the surface of the apparent horizon
and $r_{\theta}\equiv \partial r/\partial \theta$. This is the
elliptic partial differential equation about $\theta$ and $\phi$. 
It is difficult in general to solve  numerically Eq.(\ref{ah}) 
efficiently and robustly, and
several methods have been  proposed so far\cite{nko,bcsst,3d,gun}. 

Since we are interested in the criterion of the formation of horizons, 
it is important to find horizons robustly. In axisymmetric spaces, 
Eq.(\ref{ah}) can be solved as a two-point boundary value problem 
\cite{sasaki}.  And it is proved to be a robust method to find apparent
horizons since the regularity of
the apparent horizon on the symmetry axis is 
automatically guaranteed. It is desirable to
solve Eq.(\ref{ah}) in a similar manner in the non-axisymmetric space. 
To do so, we assume that the space has a discrete symmetry such that 
$r(\theta,\phi +\pi)=r(\theta,\phi)$. We also assume the space has a
reflection symmetry about $\theta=\pi/2$ plane. The latter
assumption is just for simplicity and is not essential for the method.  
Then the axis $\theta=0$ becomes the symmetry axis as in the
axisymmetric case. Thus we can specify the boundary condition at
$\theta=0,\pi/2$ as
\beq
r_{\theta}=0.
\label{bc}
\eeq
Hence we can solve Eq.(\ref{ah}) as a boundary value
problem\cite{3d} although the space is not axisymmetric. 
By finite-differencing Eq.(\ref{ah}) with taking account of the discrete
symmetry and the boundary condition Eq.(\ref{bc}), it becomes a matrix
equation, and we solve it iteratively. 
The detailed numerical 
implementation of the method is given in \cite{3d}. 
When the apparent horizon forms, we compute its
area $A$ 
\beq
A=4\int_{0}^{\pi}\int_{0}^{\pi/2}\psi^4r^2\sin\theta d\phi d\theta
\sqrt{1+{r_{\theta}^2\over r^2}+{r_{\phi}^2\over r^2\sin^2\theta}}.
\eeq
In the numerical results shown below, we discretize $\theta$ and
$\phi$ as follows
\beq
\theta_{i}=\left(i-{1\over 2}\right){\pi\over 2N_{\theta}},
~~\phi_{j}=\left(j-{1\over 2}\right){\pi \over N_{\phi}}, 
\eeq
with $i=1,\dots, N_{\theta}$, and $j=1,\dots, N_{\phi}$.  
We typically take the grid numbers  $N_{\theta}=N_{\phi}=64$ and
sometimes $N_{\theta}=N_{\phi}=128$ when a higher resolution is
required. 

\subsection{Calculating Hoop}

To gauge the hoop conjecture, we calculate the proper lengths of
geodesics of various orientations that enclose the entire
configuration. 
Due to the symmetry such that  the reflection symmetry about the
equatorial plane and the discrete symmetry about the $\theta=0$ axis, 
the equatorial ($\theta=\pi/2$) plane and some 
$\phi=$constant surfaces take the special position.  The appropriate 
circumference should be the maximum of (i) the minimum equatorial
circumference $\C^{min}_{eq}$ and (ii) the minimum polar circumference 
along $\phi=$ constant plane $\C^{min}_{pol}$. 

The geodesic equation on the equatorial plane is given by 
\beq
r_{\phi\phi}=-2 {\psi_{\phi}\over \psi}{r_{\phi}^3\over r^2} 
+\left( 2{\psi_r\over \psi}+{2\over
r}\right)r_{\phi}^2
-2{\psi_{\phi}\over \psi} r_{\phi}+ 
\left(2{\psi_{r}\over \psi} +{1\over r}\right)r^2.
\label{geo:eq}
\eeq
We note that $\C^{min}_{eq}$ is well-defined only if the equatorial 
plane symmetry is assumed as in the recent 3D dynamical calculations 
on coalescing binary NSs/BHs\cite{binary}. 
On the other hand,  geodesics along the $\phi=$ constant plane
do not exist in general. However, if we restrict ourselves to 
the plane with the symmetry such that $\psi_{\phi}=0$, then such geodesics do 
exist and follow the equation: 
\beq
r_{\theta\theta}=-2 {\psi_{\theta}\over \psi}{r_{\theta}^3\over r^2} 
+\left( 2{\psi_r\over \psi}+{2\over
r}\right)r_{\theta}^2
-2{\psi_{\theta}\over \psi} r_{\theta}+ 
\left(2{\psi_{r}\over \psi} +{1\over r}\right)r^2.
\label{geo:pol}
\eeq
Eq.(\ref{geo:eq}) and
Eq.(\ref{geo:pol}) are the same 
except for the replacement $\phi \rightarrow \theta$, 
and we solve them by the fourth-order Runge-Kutta method. 
We vary $r$ and $r_{\phi}(r_{\theta})$ at $\phi=0(\theta=0)$ 
and compute ${\cal C}$ encompassing all black holes using
Eq.(\ref{geo:eq})(Eq.(\ref{geo:pol})) up to
$\phi=\pi(\theta=\pi/2)$. In this way we obtain the minimum
circumference:
\beqa
\C^{min}_{eq}&=&2\int^{\pi}_0\psi^2\left(r^2+r_{\phi}^2\right)^{{1\over
2}}d\phi,\\
\C^{min}_{pol}&=&4 \int^{\pi/2}_0\psi^2\left(r^2+
r_{\theta}^2\right)^{{1\over
2}}d\theta.
\eeqa
We note that when we compute the minimum 
polar circumference, we first minimize the circumference with being $\phi$
fixed and then maximize over $\phi$ which satisfies $\psi_{\phi}=0$. 

\subsection{Numerical Results}

Now we construct the vacuum initial data.  We consider the
time-symmetric initial data, that is, $K_{ij}=0$. 
Then the only constraint equation we have to solve is the Hamiltonian
constraint equation
\beq
\Delta \psi=0,
\label{hamilt}
\eeq
where $\Delta$ denotes the flat Laplacian. We consider the following
configurations: (1) black holes distributed on a ring, (2) black holes 
on a spherical surface.

\subsubsection{Case (1)} 

We show the numerical results for the case (1):  black holes
distributed on a ring. 

We consider $N$ black holes (where $N$ is an even number) of an equal mass 
distributed on a ring of radius $a$ such that each black hole's 
Cartesian coordinate is $(a\cos(2\pi l/N),a\sin(2\pi l/N),0)$,
where $l=0,\dots, N-1$. Then the solution of the
constraint Eq.(\ref{hamilt}) is
\beq
\psi(r,\theta,\phi)=
1+ {M\over 2N}\sum_{l=0}^{N-1}{1\over \sqrt{r^2 -2ar \sin\theta
\cos\left(\phi-2\pi l/N\right) +a^2}},
\eeq
where $M$ is the ADM mass of the total system. In the limit of large
$N$ the configuration becomes the singular ring, and the solution
becomes
\beq
\psi(r,\theta)=1+ {M\over \pi\sqrt{r^2+2ar\sin\theta +a^2}}K(\kappa), 
\eeq
where $K(\kappa)$ is the elliptic integral of the first kind with
\beq
\kappa^2={4ar\sin\theta\over r^2+2ar\sin\theta +a^2}.
\eeq
The results are shown in Fig.1 and Table 1. 
The top view of  
the apparent horizon as well as the bird's-eye view of it is shown. 
The shape of the apparent horizon indeed respects the discrete
symmetry of the configuration. 
As $N$ increases, the shape becomes pancake-like. 
The maximum of $(\C^{min}_{eq},\C^{min}_{pol})$ is shown in the
table.  We calculate the circumference ${\C}$ of the system irrespective
of the existence of an apparent horizon. Even if the apparent horizon
forms, the appropriate circumference is generally not located on it.  
The circumference on the apparent horizon can be significantly 
larger\cite{nst,cnns,bs}.

When the apparent horizon forms, its area $A$ normalized by 
$16\pi M^2$
 is computed because in the time-symmetric initial data the following 
inequality holds if the final state is stationary:
\beq
A \leq A_{EH} \leq A_f(=16\pi M_f^2) \leq 16\pi M^2.
\eeq
Here $A_{EH}$ is the area of the event horizon. We note that the
existence of an apparent horizon implies the existence of the event
horizon enclosing it\cite{he}. From this and the fact that \
the apparent horizon is the minimum surface in the time-symmetric 
initial data\cite{gibbons}, the first inequality holds. 
$A_f$ is the final stationary state black hole's area, and the second
inequality follows from the area theorem\cite{he}. From the
uniqueness theorem\cite{israel},  the final state is the 
Schwarzschild black hole with mass $M_f$ because we are considering
the non-rotating system. $M_f$ should not be larger than the initial mass
$M$ because gravitational radiation conveys the energy during the
dynamical evolution. 

In the cases $N=2$ and $N 
\rightarrow \infty$, we also show the numerical results using a
2D(axisymmetric) apparent horizon finder\cite{sasaki} 
for comparison and the check of our numerical results. We find good
agreement. 

\subsubsection{Case (2)} 

Next we show the numerical results for the case (2):  black holes 
on a spherical surface.

Black holes are distributed on a sphere of radius $a$ such that each
black hole's Cartesian coordinate is 
\beq
(a\sin(\pi k/n)\cos(2\pi l/n),a\sin(\pi k/n)
\sin(2\pi l/n),a\cos(\pi k/n)),
\eeq
where $k=1,\dots,n-1,l=0,\dots, n-1$. We also consider two additional
black holes located at $\theta=0$, and $\pi$ on the surface 
so that in the large $N$ limit the configuration becomes a singular 
spherical surface. The total number of black holes
is then $N=n(n-1)+2$. We assume $n$ is even number. The solution of the
constraint Eq.(\ref{hamilt}) is simply
\beqa
\psi(r,\theta,\phi)&=&1+ {M\over 2N}\left({1\over 
\sqrt{r^2+2ar\cos\theta+a^2}}+{1\over
\sqrt{r^2-2ar\cos\theta+a^2}}
\right)\nonumber\\
&+&{M\over 2N}\sum_{k=1}^{n-1}\sum_{l=0}^{n-1}{1\over
\sqrt{r^2+a^2-2ar \left( \sin\theta\sin(\pi k/n)\cos(\phi-2\pi l/n)
+\cos\theta\cos(\pi k/n)\right)}}.
\eeqa

The results are shown in Fig.2 and Table 2. For smaller $N$ 
the apparent horizon is  bumpy. 
Since the system becomes spherically symmetric in the large $N$ limit, 
$A/16\pi M^2$ becomes close to 1 as $N$ increases and $\C$ does
not clearly distinguish the existence or non-existence of the apparent 
horizon.

\section{summary}

We tested the hoop conjecture in spaces without axisymmetry but with a 
discrete symmetry. The existence or nonexistence of an apparent
horizon encompassing all black holes is qualitatively consistent with
the hoop conjecture proposed by Thorne. From Table 1 and Table 2,  we
find that if the  circumference $\C$ satisfies
$\C/4\pi M \siml 1.168$, then the configuration will be
surrounded by a common apparent horizon. 

In non-axisymmetric spaces with a discrete symmetry,  
searching for hoops 
is less computationally demanding than searching for apparent 
horizons: just solving the ordinary differential equation. 
Further if the equatorial plane symmetry is assumed
as in the recent 3D dynamical calculations on 
coalescing binary NSs/BHs\cite{binary}, 
$\C_{eq}^{min}$ is then well-defined, and therefore even in 
more general spaces the hoop concept can be a useful diagnostic for
the final fate of the collapsed object.

\acknowledgments
The author would like to thank Professor T.Nakamura for encouragement
and Professor N.Sugiyama for a careful reading of the manuscript. 
He is also grateful to an anonymous referee for useful comments which
improved the manuscript. 
This work was supported in part by YITP and in part by a JSPS 
Fellowship for Young Scientists under grant No.3596.

\newpage

\vskip 0.3in
\centerline{FIGURE CAPTION}
\vskip 0.05in

\newcounter{fignum}
\begin{list}{Fig.\arabic{fignum}.}{\usecounter{fignum}}

\item
The shape of the apparent horizon for the case (1) 
near the critical separation is shown 
for (a) $N=2$, (b) $N=4$, (d) $N=6$, (d) $N=10$ and 
(e) $N=\infty$. 
The top view of the apparent horizon 
as well as the bird's-eye view of it is shown. 
As $N$ increase the shape becomes pancake-like. The coordinates are 
in units of $M$.

\item
The shape of the apparent horizon for the case (2) 
near the critical separation is shown 
for (a) $N=4$, (b) $N=14$, (c) $N=32$ and (d) $N=58$. 
For smaller $N$ the surface  is bumpy. 
As $N$ increases it becomes spherical.

\end{list}

\vskip 0.3in
\centerline{TABLE CAPTION}
\vskip 0.05in

\newcounter{tabnum}
\begin{list}{Table \arabic{tabnum}.}{\usecounter{tabnum}}

\item
Properties of black holes distributed on a ring. Here ``2D'' 
indicates the numerical results using an  axisymmetric apparent horizon
finder. ``Y'' means that apparent horizon enclosing all black
holes are found; ``N'' means the
opposite. $\C=\max(\C^{min}_{eq},\C^{min}_{pol})$.

\item
Properties of black holes  on a spherical surface.  

\end{list}

\newpage

{\bf Table 1a:}
N=2 case

\vspace{0.5cm}

\begin{tabular}{ccccc} \hline
$a/M$ & AH? &  $\C/4\pi M$ & $A/16\pi M^2$ 
& $A/16\pi M^2$(2D) \\ \hline
0.38  &  Y      &        1.114 & 0.9780  & 0.9780 \\
0.383 & Y       &        1.115 &  0.9769 & 0.9769\\ 
0.39  & N     &  1.118 & & \\ \hline
\end{tabular}

\vspace{1.5cm}

{\bf Table 1b:}
N=4 case

\vspace{0.5cm}

\begin{tabular}{cccc} \hline
$a/M$ & AH? &  $\C/4\pi M$ & $A/16\pi M^2$  \\ \hline
0.41 &  Y    &          1.140 &  0.9865  \\
0.42 & Y     &  1.144  &  0.9841 \\     
0.43  & N     &  1.149 &   \\ \hline
\end{tabular}

\vspace{1.5cm}

{\bf Table 1c:}
N=6 case

\vspace{0.5cm}

\begin{tabular}{cccc} \hline
$a/M$ & AH? &  $\C/4\pi M$ & $A/16\pi M^2$  \\ \hline
0.43  &  Y      &   1.154 & 0.9883   \\
0.44 &  Y    &      1.159 &  0.9860  \\
0.45 &  N     &  1.163 &   \\ \hline
\end{tabular}

\vspace{1.5cm}

{\bf Table 1d:}
N=10 case

\vspace{0.5cm}

\begin{tabular}{cccc} \hline
$a/M$ & AH? &  $\C/4\pi M$ & $A/16\pi M^2$  \\ \hline
0.45  &  Y      &         1.163 & 0.9875   \\
0.46 &  Y    &          1.168 &  0.9857  \\
0.47 & N     &  1.173 &   \\ \hline
\end{tabular}

\vspace{1.5cm}

{\bf Table 1e:}
N=$\infty$ (ring) case

\vspace{0.5cm}

\begin{tabular}{ccccc} \hline
$a/M$ & AH? &  $\C/4\pi M$ & $A/16\pi M^2$ 
& $A/16\pi M^2$(2D) \\ \hline
0.49  &  Y     &  1.155 &  0.9817  &   0.9812    \\
0.495 &  Y     &  1.158 &  0.9794  &  0.9797 \\
0.505 & N      & 1.163      &  & \\ \hline
\end{tabular}

\vspace{7cm}

{\bf Table 2a:}
N=4  case
\vspace{0.5cm}

\begin{tabular}{cccc} \hline
$a/M$ & AH? &  $\C/4\pi M$ & $A/16\pi M^2$  \\ \hline
0.40  &  Y      &  1.136 & 0.9890   \\
0.41 &  Y    &    1.140 &  0.9870  \\
0.42 & N     &  1.145  &       \\  \hline
\end{tabular}

\vspace{1.5cm}

{\bf Table 2b:}
N=14  case
\vspace{0.5cm}

\begin{tabular}{cccc} \hline
$a/M$ & AH? &  $\C/4\pi M$ & $A/16\pi M^2$  \\ \hline
0.40  &  Y     &  1.068  &  0.9982 \\
0.43  &  Y     &  1.076 & 0.9965   \\
0.44 &  Y    &    1.079 &  0.9954  \\
0.45 & N     &  1.082 &    \\   \hline
\end{tabular}

\vspace{1.5cm}

{\bf Table 2c:}
N=32  case
\vspace{0.5cm}

\begin{tabular}{cccc} \hline
$a/M$ & AH? &  $\C/4\pi M$ & $A/16\pi M^2$  \\ \hline
0.40 & Y     &  1.054   &  0.9992  \\
0.44 &  Y    &  1.062  &  0.9982  \\
0.45 & Y     &  1.064  &  0.9975    \\
0.46 & N     &  1.066  &             \\   \hline
\end{tabular}

\vspace{1.5cm}

{\bf Table 2d:}
N=58  case
\vspace{0.5cm}

\begin{tabular}{cccc} \hline
$a/M$ & AH? &  $\C/4\pi M$ & $A/16\pi M^2$  \\ \hline
0.40 & Y     & 1.051   &  0.9993  \\
0.44 &  Y    &   1.058 &  0.9986  \\
0.45 & Y     &  1.060  &  0.9983 \\   
0.46 & N     &  1.061  &   \\ \hline
\end{tabular}

\vspace{1.5cm}  

\newpage

\begin{figure}
\centering
\epsfig{figure=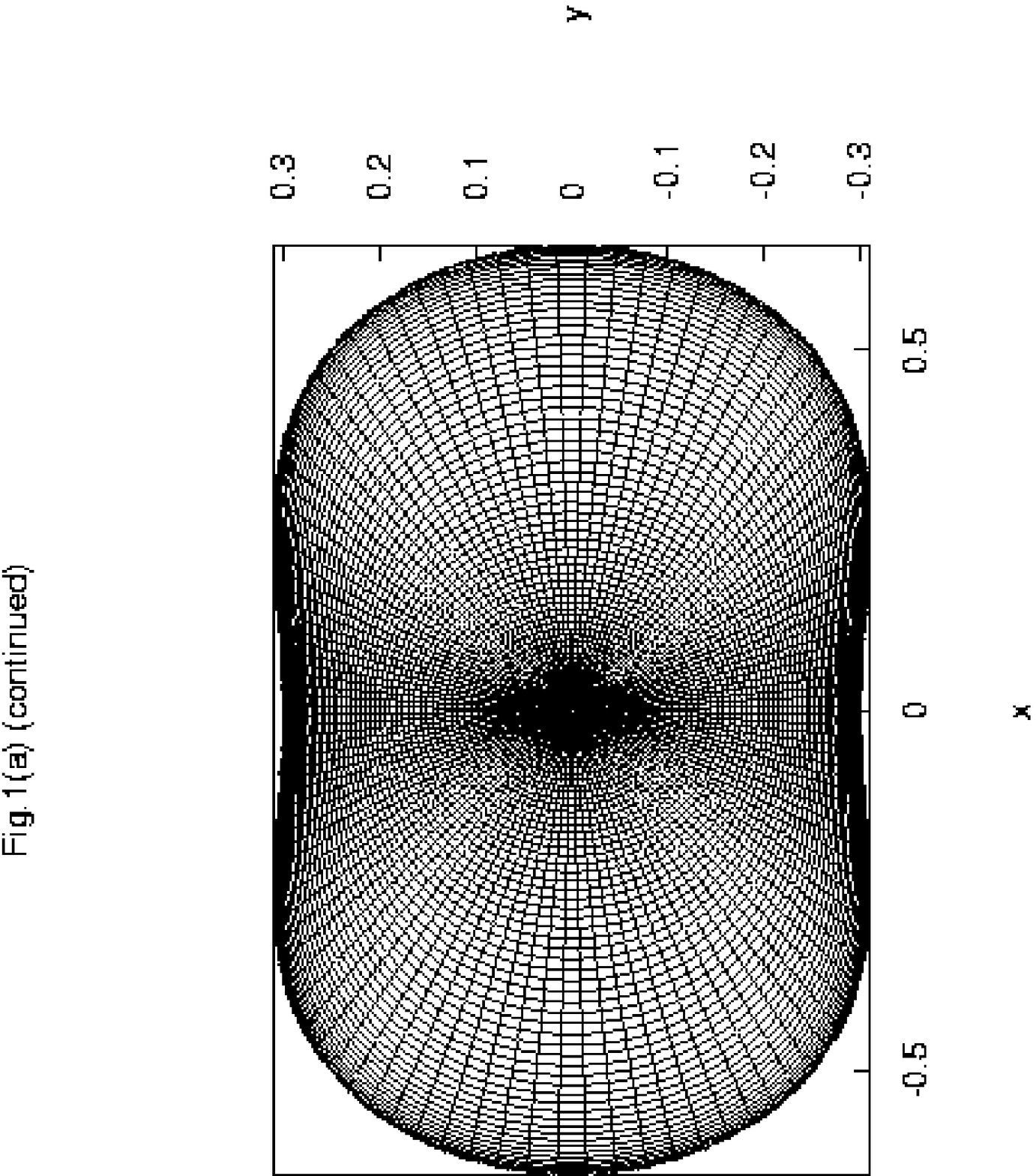, width=8cm}
\end{figure}
\begin{figure}
\centering
\epsfig{figure=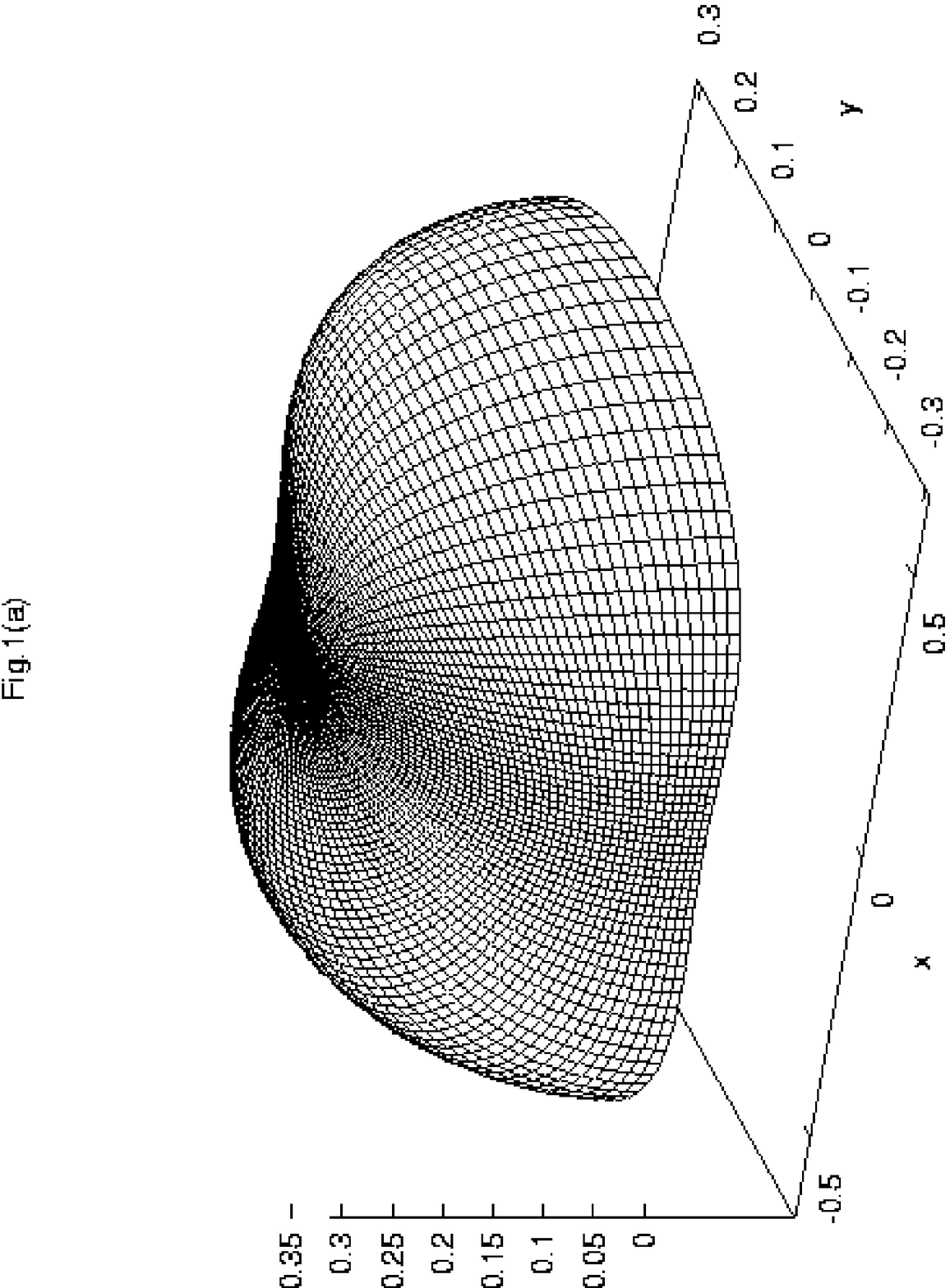, width=8cm}
\end{figure}

\begin{figure}
\centering
\epsfig{figure=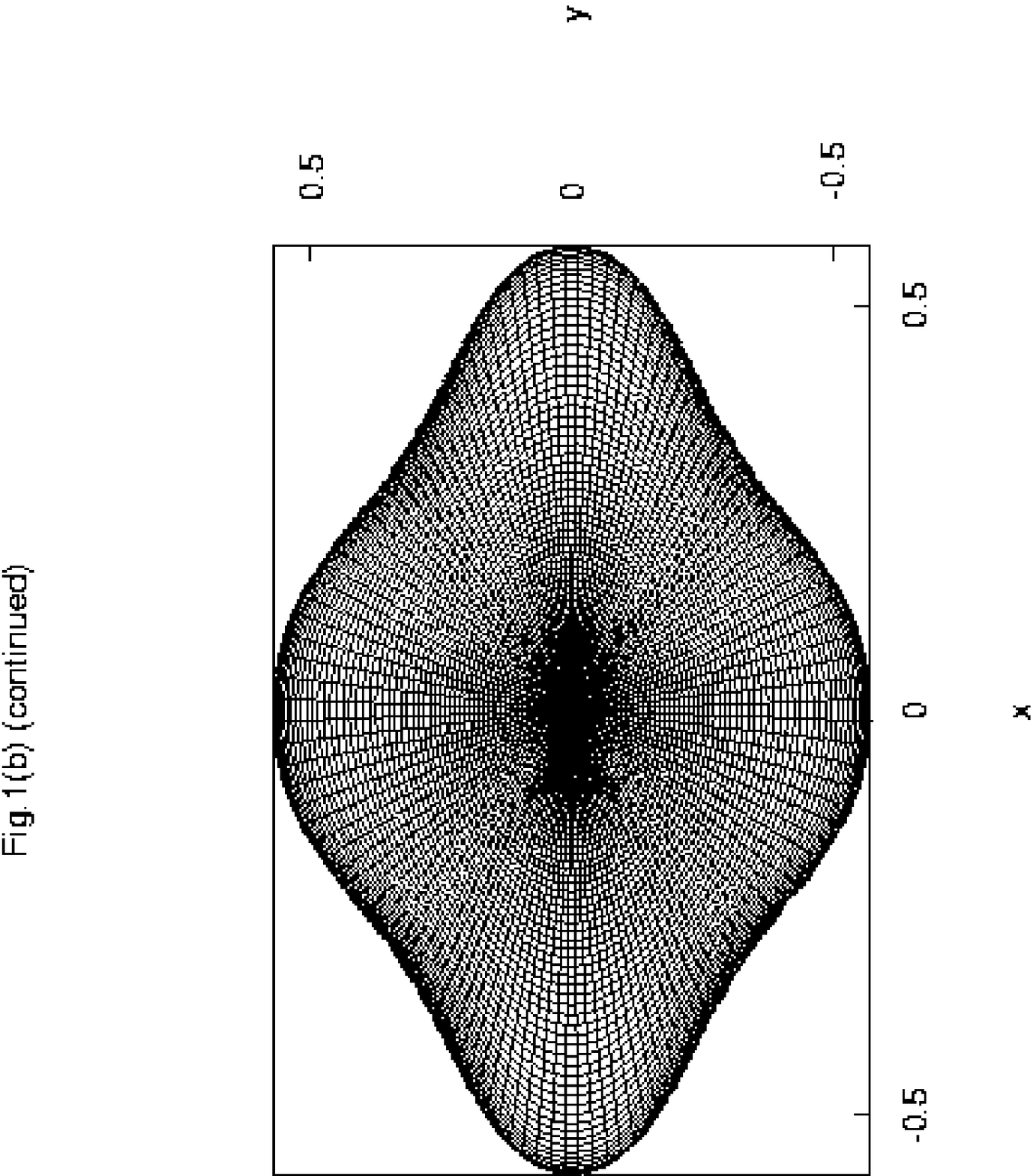, width=8cm}
\end{figure}
\begin{figure}
\centering
\epsfig{figure=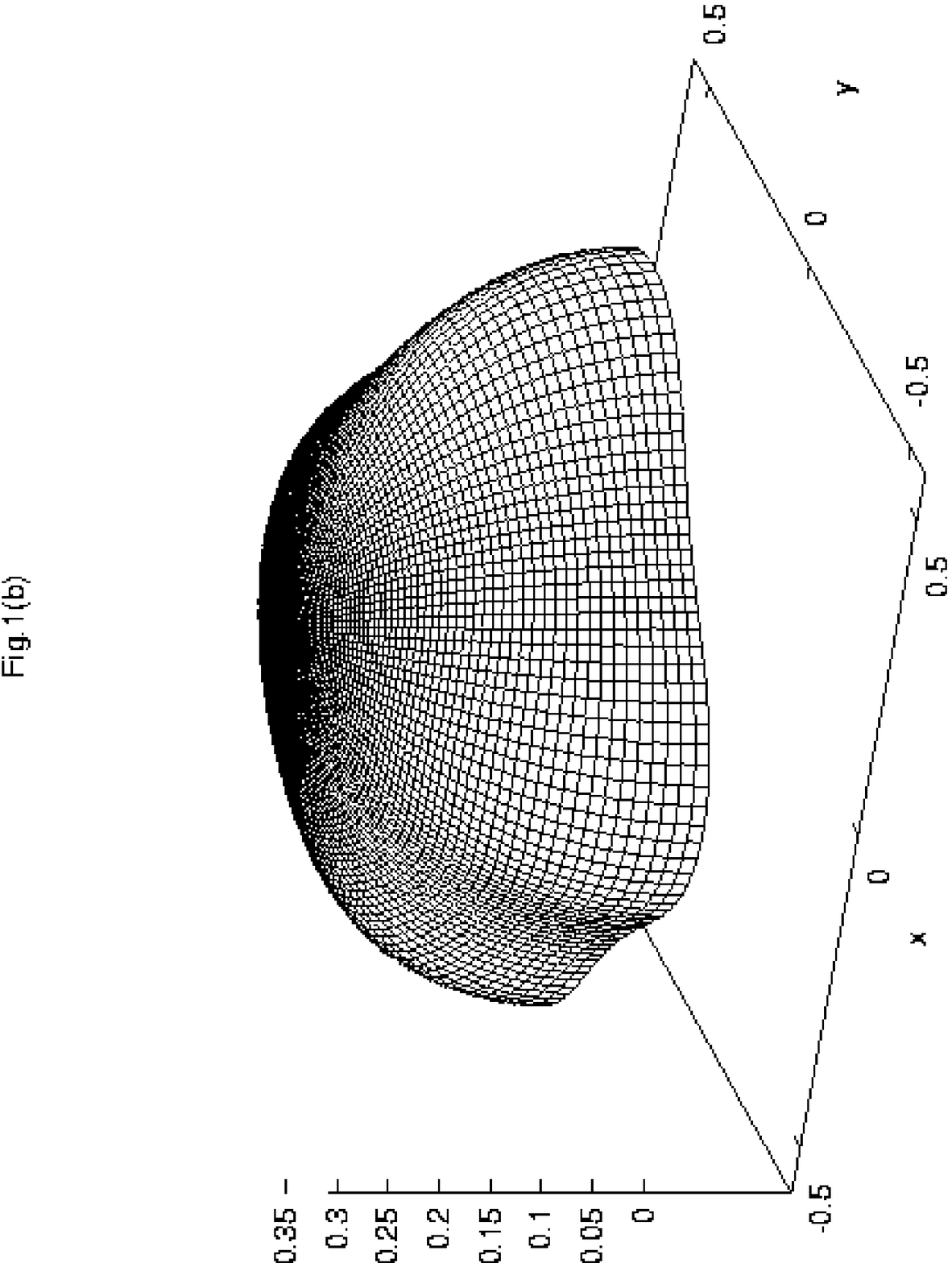, width=8cm}
\end{figure}

\begin{figure}
\centering
\epsfig{figure=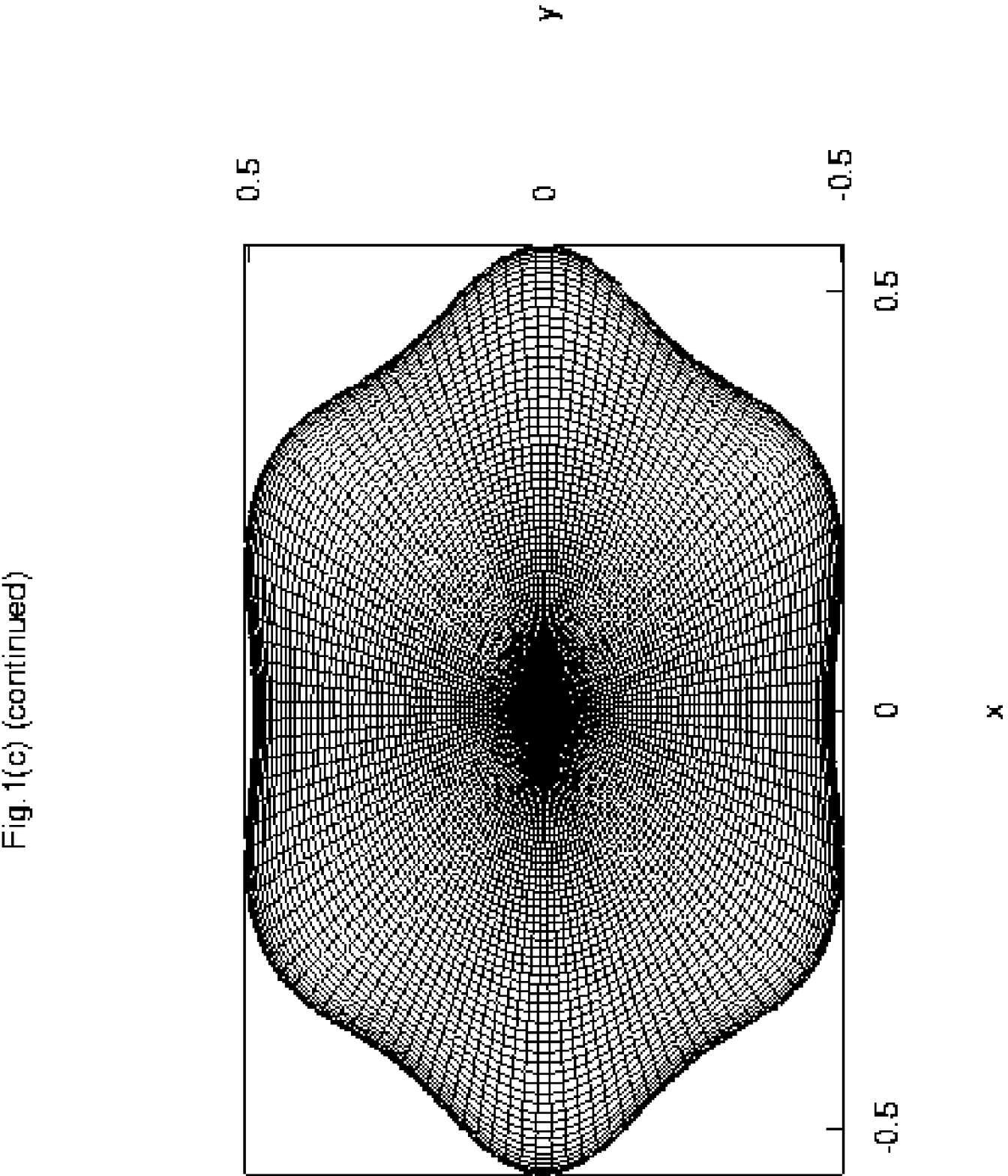, width=8cm}
\end{figure}
\begin{figure}
\centering
\epsfig{figure=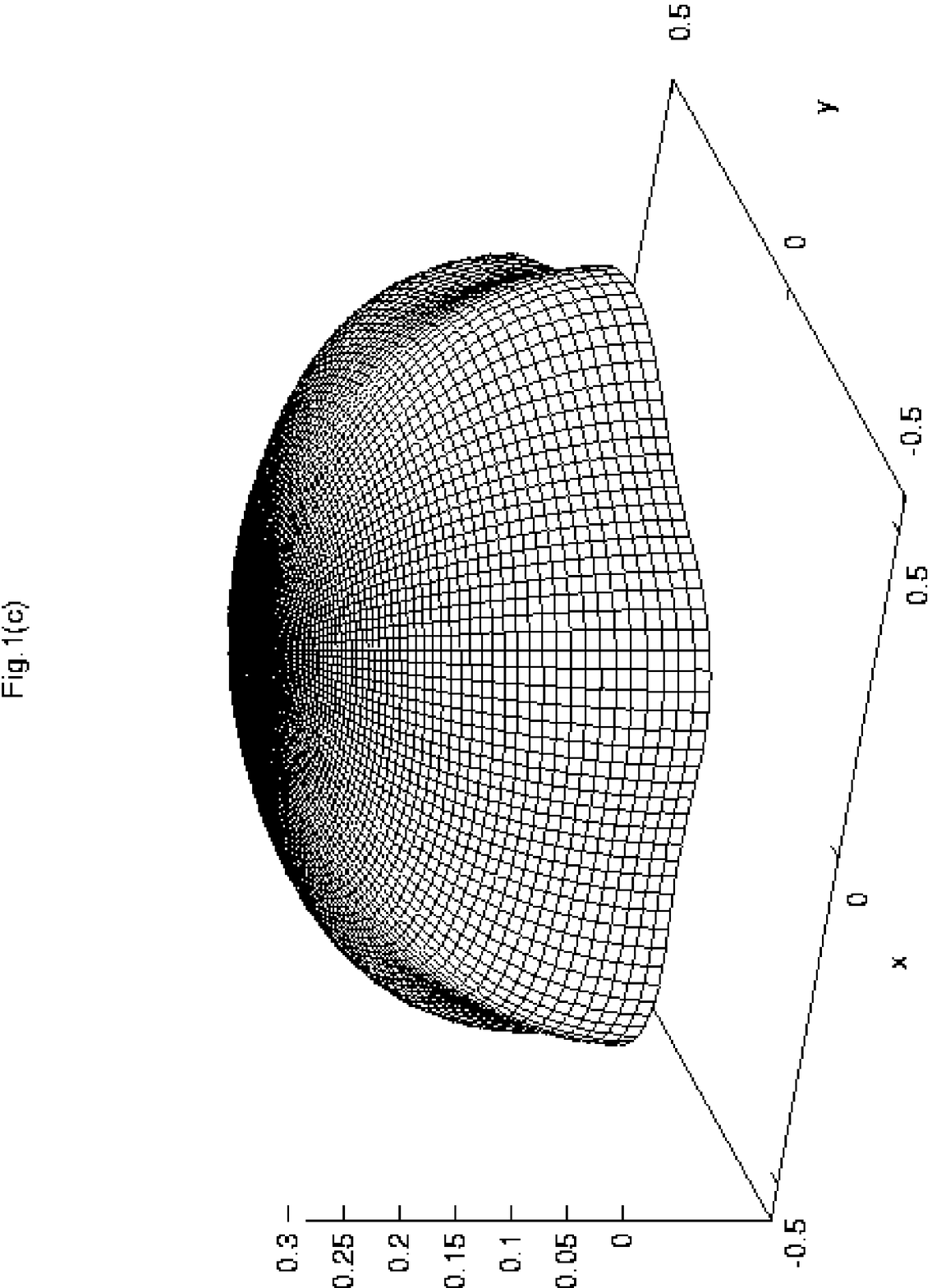, width=8cm}
\end{figure}

\begin{figure}
\centering
\epsfig{figure=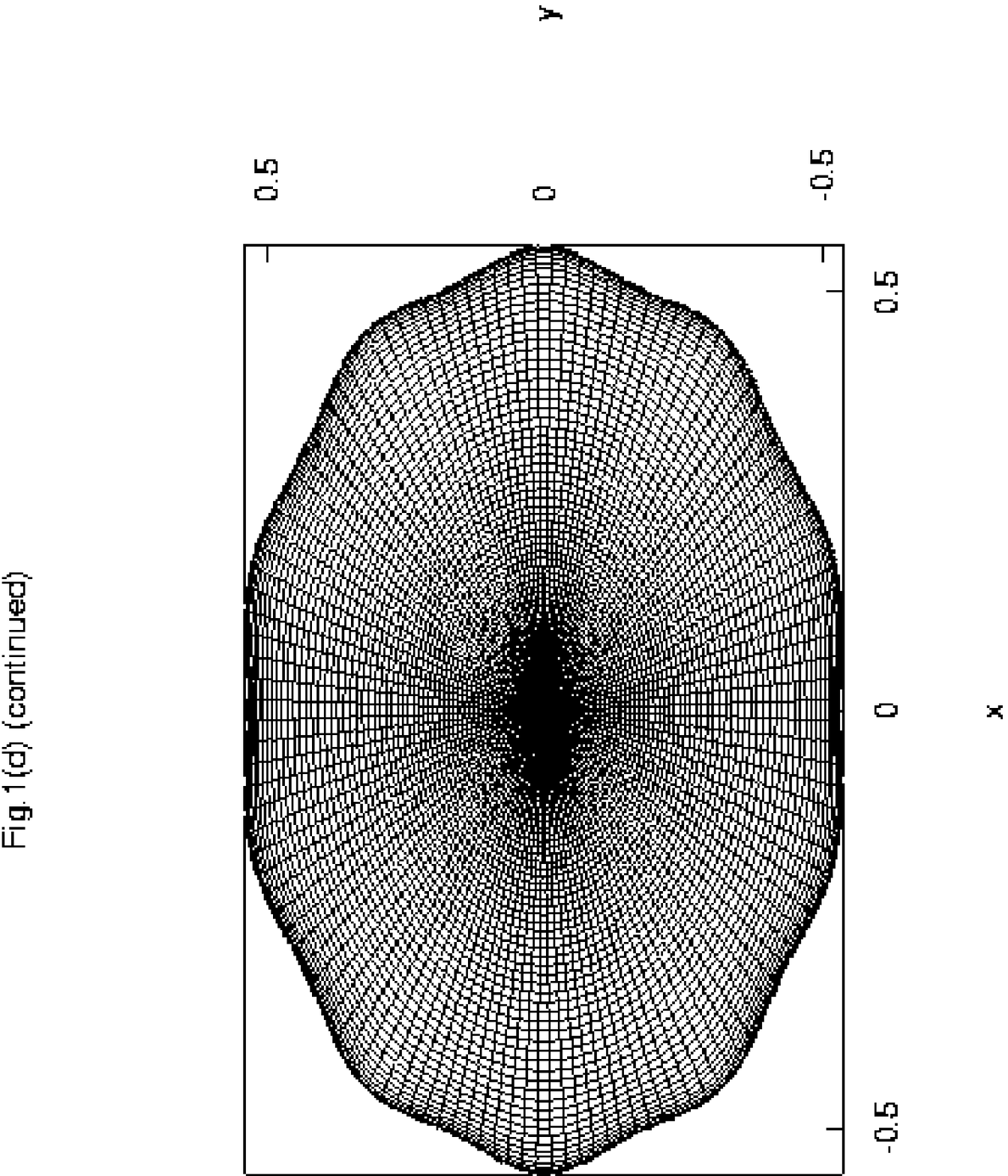, width=8cm}
\end{figure}
\begin{figure}
\centering
\epsfig{figure=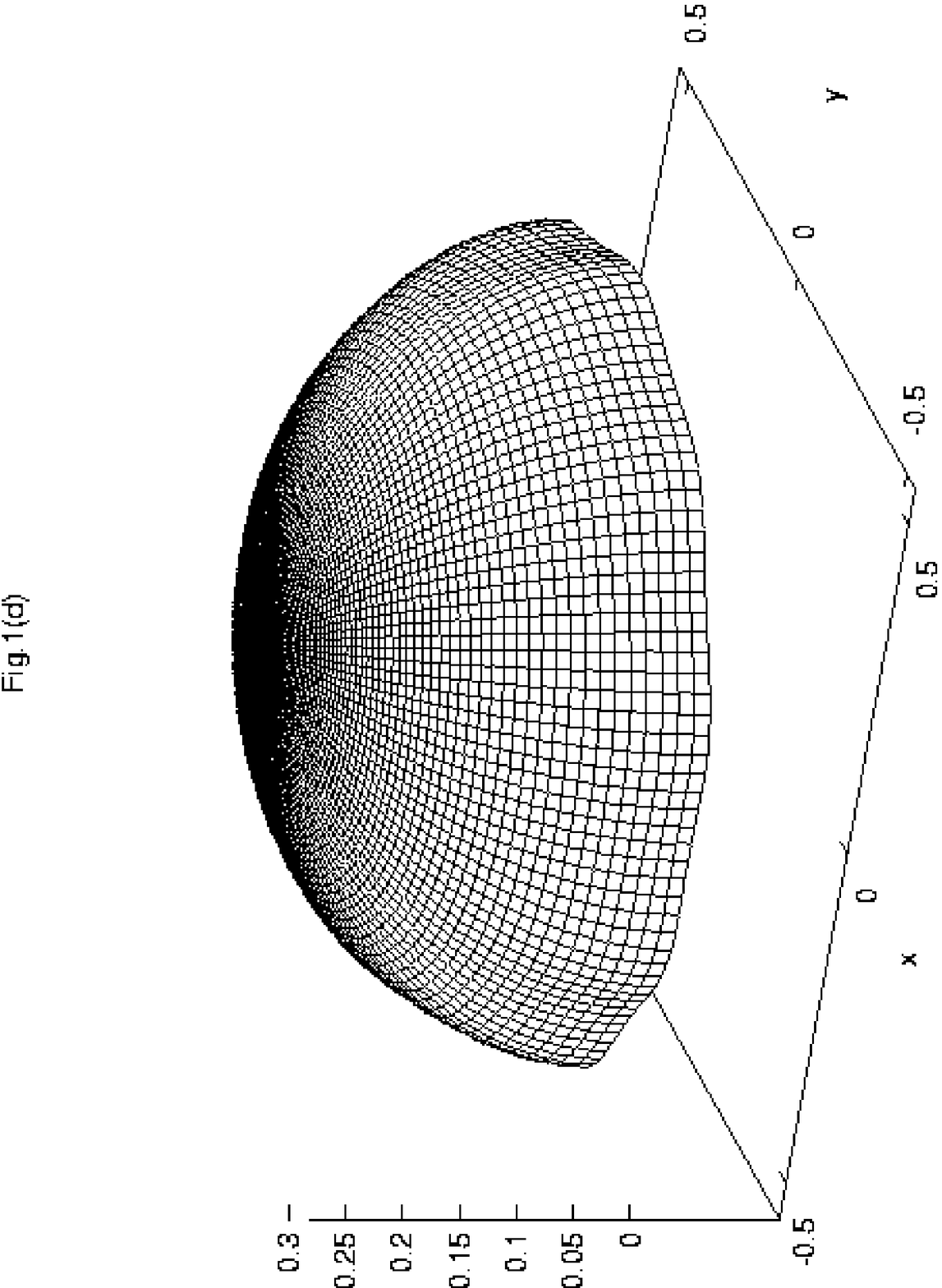, width=8cm}
\end{figure}

\begin{figure}
\centering
\epsfig{figure=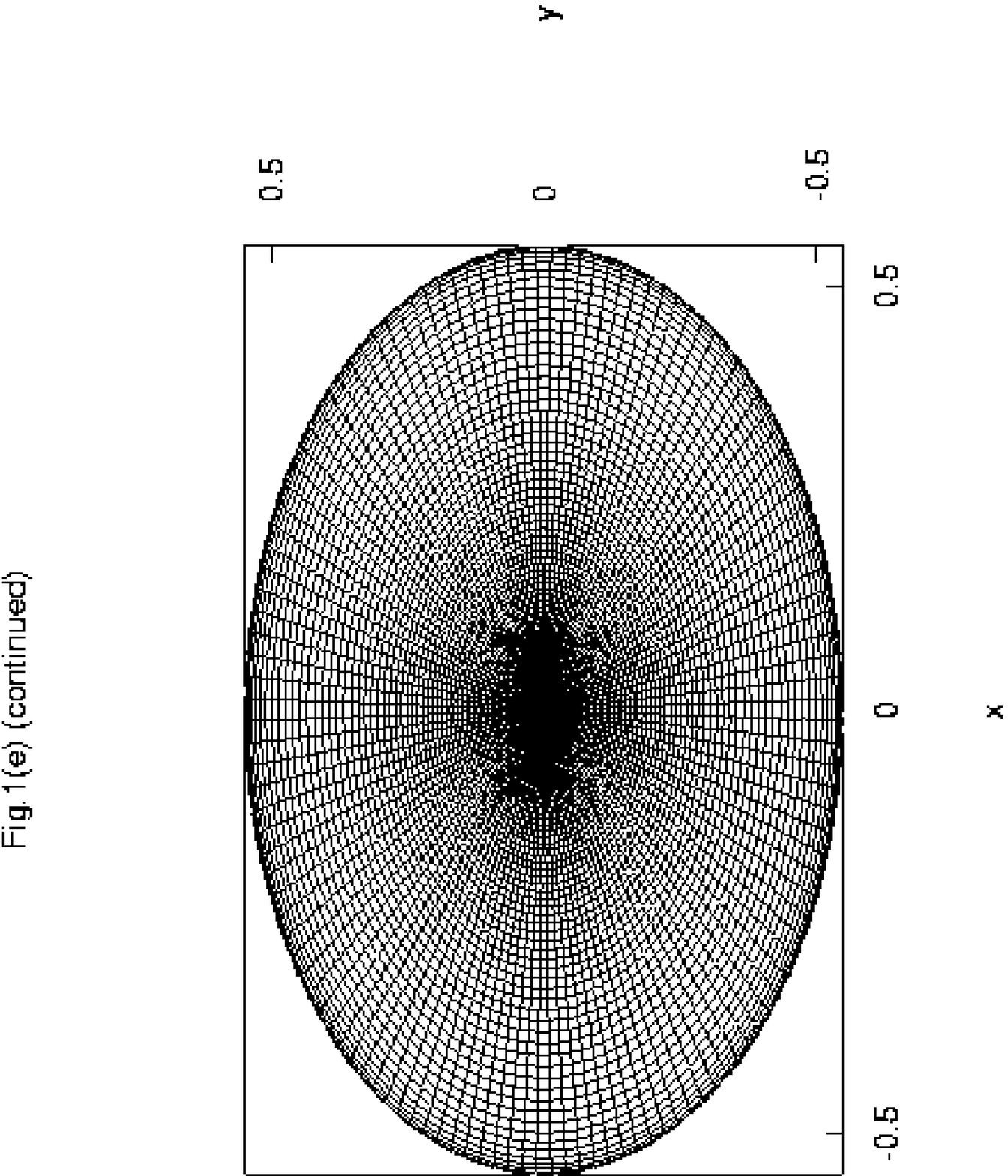, width=8cm}
\end{figure}
\begin{figure}
\centering
\epsfig{figure=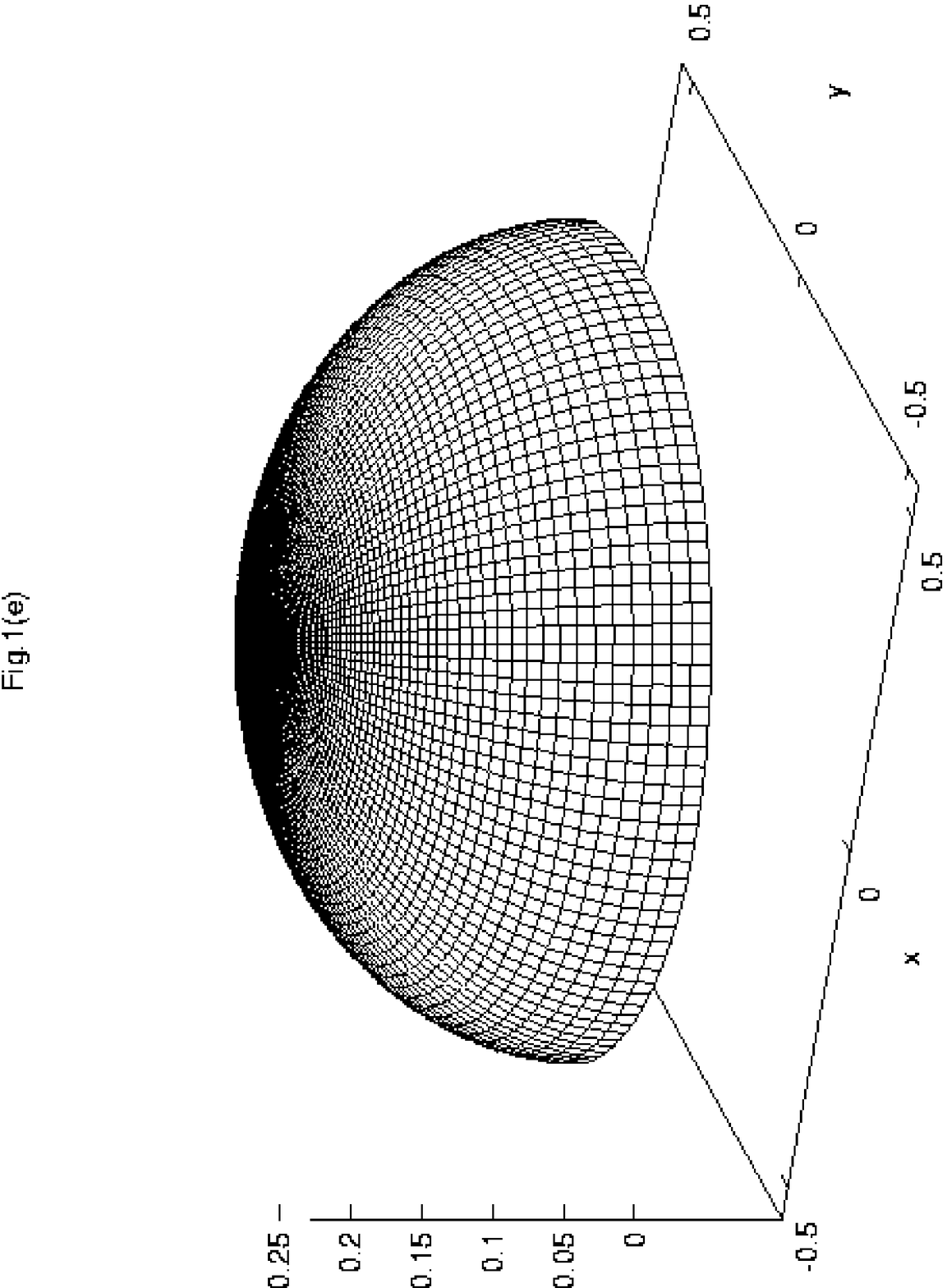, width=8cm}
\end{figure}

\begin{figure}
\centering
\epsfig{figure=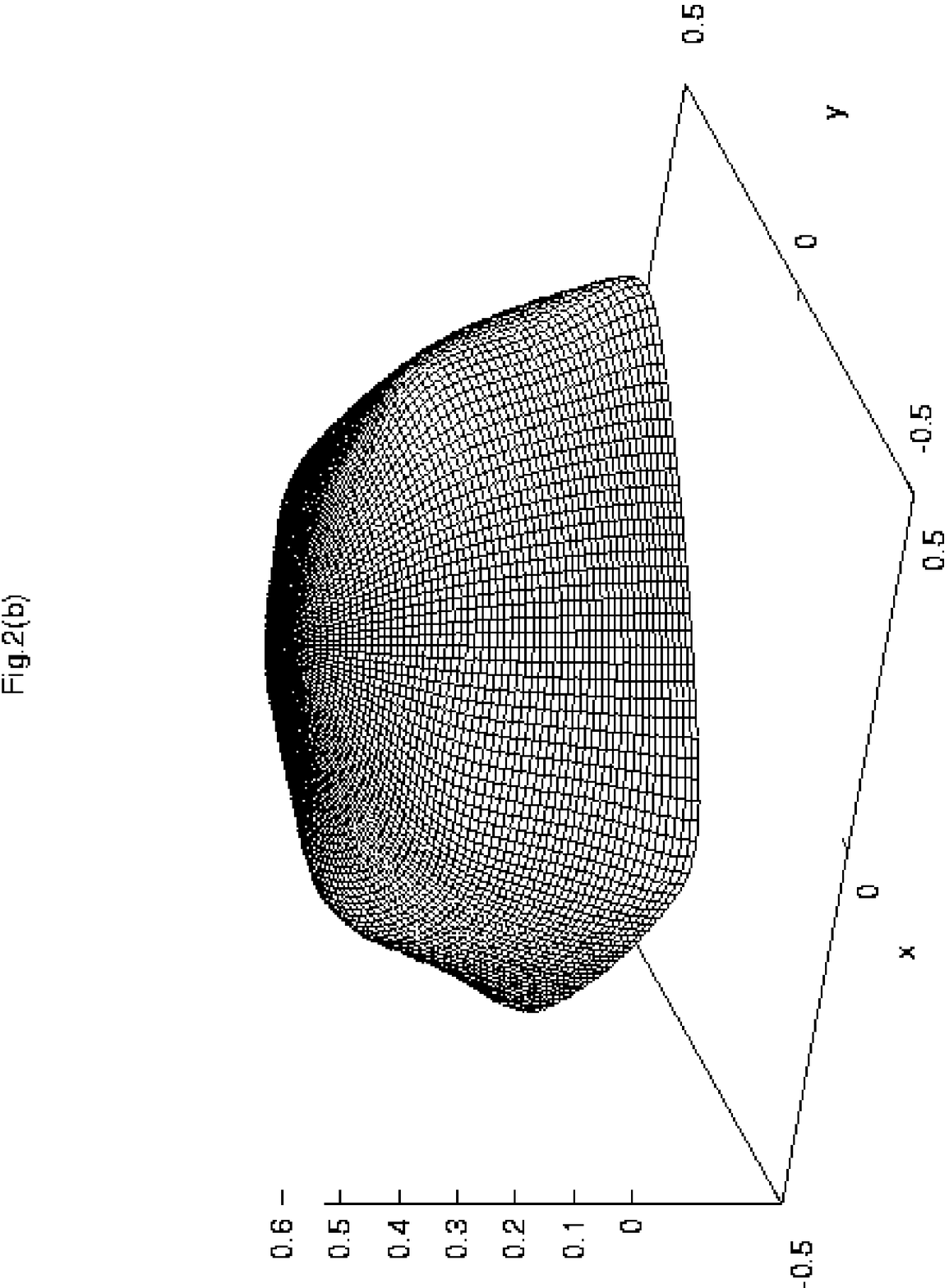, width=8cm}
\end{figure}
\begin{figure}
\centering
\epsfig{figure=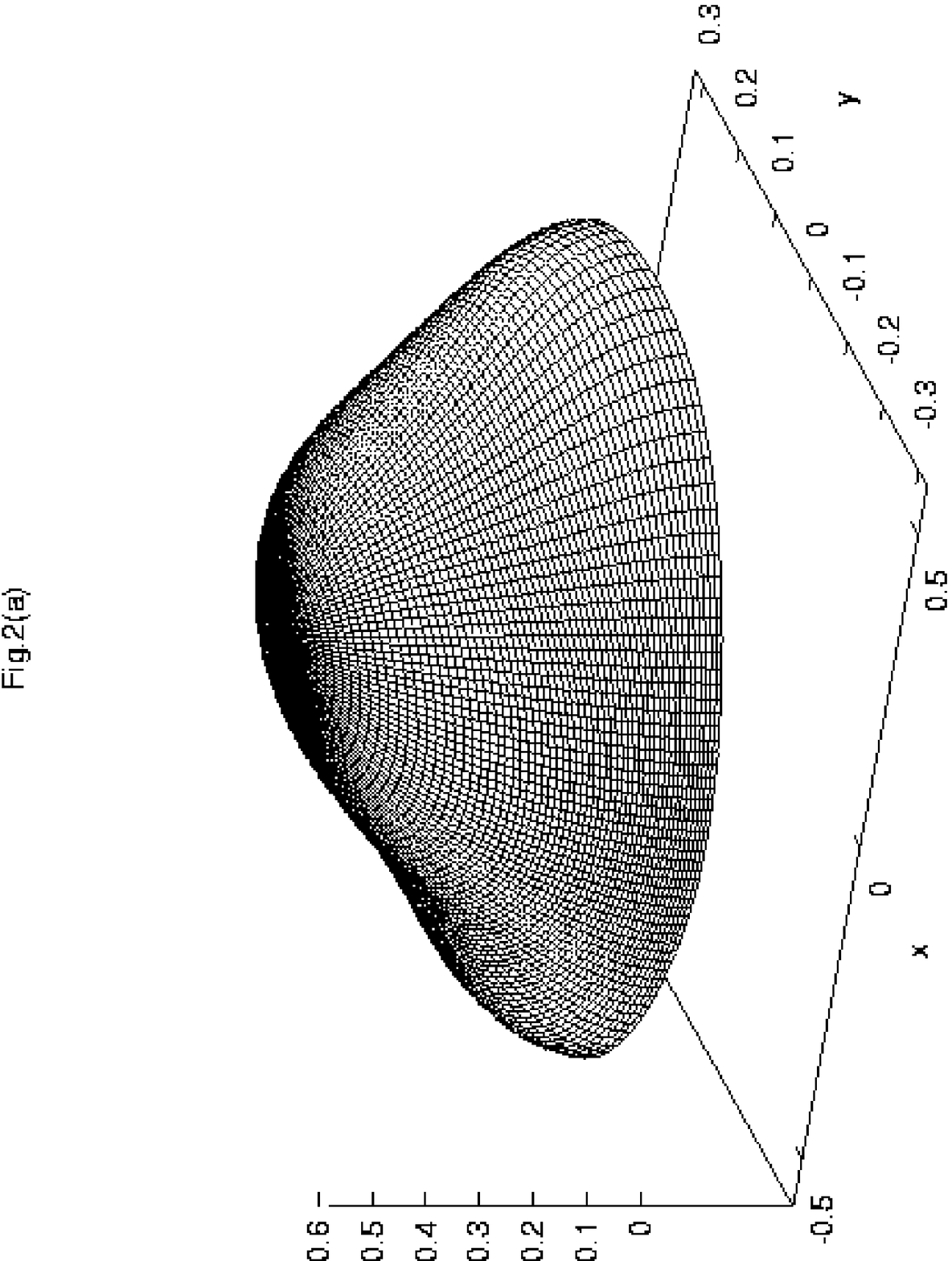, width=8cm}
\end{figure}
\begin{figure}
\centering
\epsfig{figure=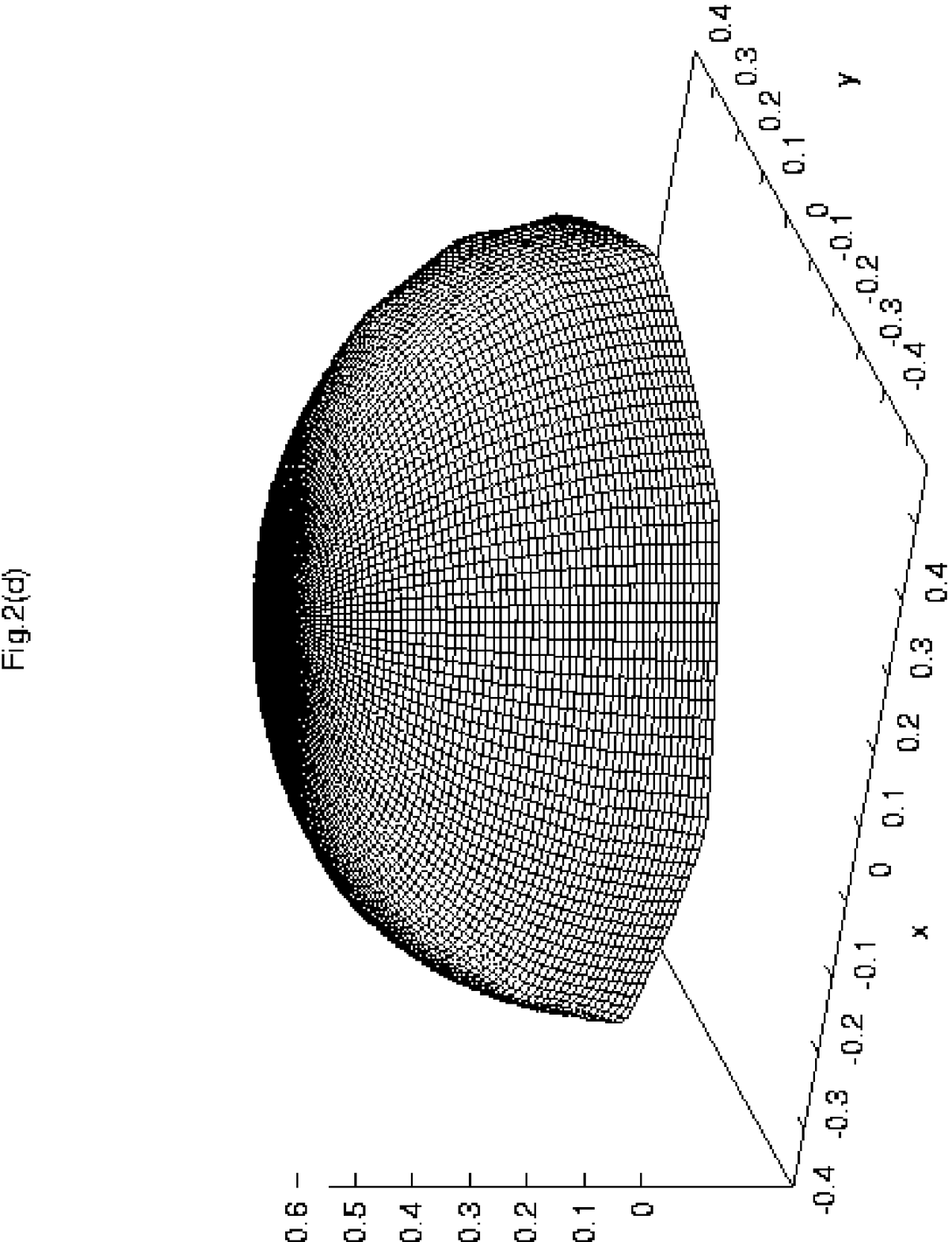, width=8cm}
\end{figure}
\begin{figure}
\centering
\epsfig{figure=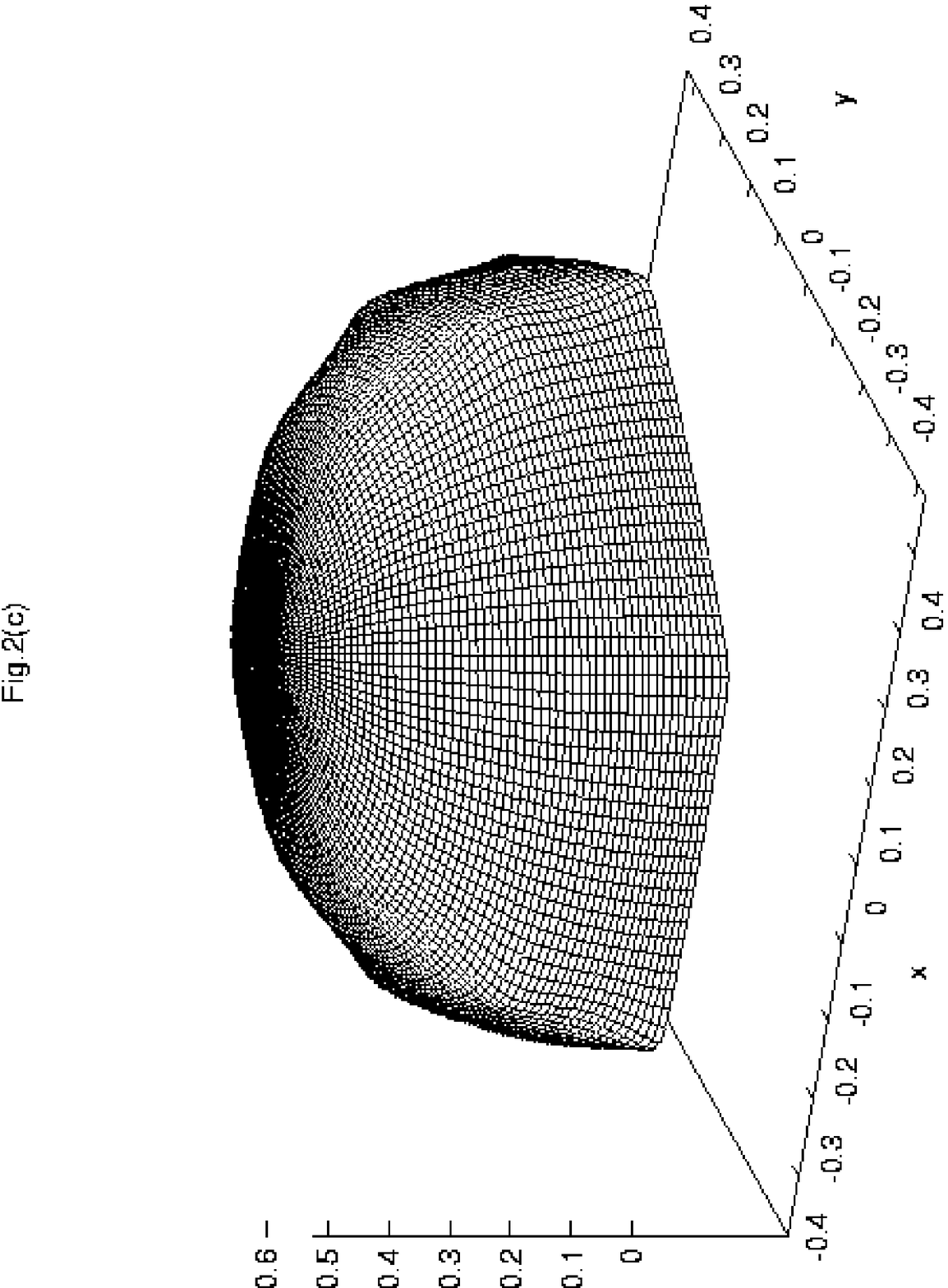, width=8cm}
\end{figure}

\end{document}